\documentclass{rspublic}
\usepackage{natbib}
\usepackage{amsmath}
\usepackage{amsfonts}
\newcommand{\tno}{\ensuremath{{2n+1}}}

\usepackage{latexsym}
\usepackage{amsmath}
\usepackage{amssymb}
\usepackage{amsfonts}
\usepackage{ifthen}

\def\01{\{0,1\}}

\newcommand{\eps}{\varepsilon}

\newtheorem{definition}{Definition}

\newcounter{protoCount}
\newcounter{protoList}
\newsavebox{\tmpbox}
\newlength{\protobox}
\newenvironment{protocol}[2]{
\bigskip
\addtocounter{protoCount}{1}
\noindent \begin{lrbox}{\tmpbox}
\setlength{\protobox}{\textwidth}
\addtolength{\protobox}{-0.8cm}
\begin{small}
\begin{minipage}[c]{\protobox}
\begin{bfseries}Protocol \theprotoCount: #1\end{bfseries}
\ifthenelse{\equal{#2}{\empty}}{}{\\Prerequisite: #2}
\begin{list}{\begin{bfseries}\arabic{protoList}:\end{bfseries}}
{\usecounter{protoList}}
}{ 
\end{list}
\end{minipage}
\end{small}
\end{lrbox}
\fbox{\usebox{\tmpbox}}
\bigskip
}  
\begin{document}
\title[Implications of Superstrong Nonlocality for Cryptography]{Implications of Superstrong Nonlocality for Cryptography}
\author[H. Buhrman, M. Christandl, F. Unger, S. Wehner and A. Winter]
{Harry Buhrman$^{1,2}$, Matthias Christandl$^{3}$, Falk Unger$^{2}$,\\ Stephanie Wehner$^{2}$ and Andreas Winter$^{4}$}
\affiliation{
$^{1}$ University of Amsterdam\\
$^{2}$ Centrum voor Wiskunde en Informatica,
Kruislaan 413,\\ 1098 SJ Amsterdam, The Netherlands\\
$^{3}$ Centre for Quantum Computation, Department of Applied Mathematics and
Theoretical Physics, University of Cambridge, Wilberforce Road,
\\Cambridge CB3 0WA, United Kingdom\\
$^{4}$ Department of Mathematics, University of Bristol,
University Walk,\\ Bristol BS8 1TW, United Kingdom
}
\label{firstpage}
\maketitle

\begin{abstract}{nonlocality, cryptography}
Non-local boxes are hypothetical ``machines'' that give rise to superstrong
non-local correlations, leading to a stronger violation of Bell/CHSH
inequalities than is possible within the framework of quantum mechanics.
We show how non-local boxes can be used to perform any two-party
secure computation. We first construct a protocol for bit commitment and
then show how to achieve oblivious transfer using non-local boxes. Both
have been shown to be impossible using quantum mechanics alone. 
\end{abstract}

\section{Introduction}

Consider two parties, Alice (A) and Bob (B), who are not able to
communicate but have access to physical states that they can use
to generate joint correlations. The generation of correlation
can be regarded as an experiment in which both parties decide to
measure the state of their system, and the outcomes of their
measurements are given by random variables. Classical as well
as quantum theories put limits on non-local correlations
that can be generated between separated sites when no
communication is available. In particular, both classical and
quantum theories do not violate the no-signaling condition of
special relativity, i.e. the local choice of measurements
may not lead to observable differences on the other end.
The limits on the strength of correlations
generated in the framework of any classical theory (i.e. a theory
based on local hidden variables) are known as \emph{Bell
inequalities}~\citep{bell:epr}. A well-known variant of a Bell
inequality is the~\citet*{chsh:nonlocal} (CHSH)
inequality, which can be expressed as~\citep{wim:thesis}
$$
\sum_{x,y \in \01} \Pr (a_x \oplus b_y = x \cdot y) \leq 3.
$$
Here, $x \in \01$ and $y \in \01$ denote the choice of Alice's and
Bob's measurement, $a_x \in \01$ and $b_y \in \01$ the
respective binary outcomes, and $\oplus$ addition modulo $2$. 
The theory of quantum mechanics
allows the violation of this inequality, but curiously only up to
a maximal value of $2 + \sqrt{2}$ which is known as \emph{Cirel'son's
bound}~\citep{cirelson:bound}. Since special relativity
allows a violation of Cirel'son's bound, 
~\cite{popescu:nonlocal,popescu:nonlocal2,popescu:nonlocal3}
raised the question why nature is not more ``non-local''? That is,
why does quantum mechanics not allow for a stronger violation of
the CHSH inequality up to the maximal value of 4? To gain more
insight into this question, they constructed a toy-theory based on
so-called \emph{non-local boxes}. Each such box takes inputs $x,y \in \01$ from
Alice and Bob respectively and outputs measurement outcomes
$a_x$,$b_y$ such that $x\cdot y = a_x \oplus b_y$. Note that Alice
and Bob still cannot use this box to transmit any information.
However, since for all $x$ and $y$,
$\Pr(a_x \oplus b_y = x \cdot y) = 1$, the above sum equals 4 and thus
non-local boxes lead to a maximum violation of the CHSH
inequality.

In this paper, we investigate the relationship between nonlocality and
cryptography. As it has been shown~\citep{lo:insecurity,lo&chau:bitcom,lo&chau:bitcom2,mayers:trouble,mayers:bitcom},
classical as well as quantum mechanics do not allow for the
construction of unconditionally secure bit commitment and
oblivious transfer without additional
assumptions.
Thus it is a fundamental problem to assess whether \emph{any} theory
that generates correlations renders these cryptographic primitives possible,
while simultaneously preserving the no-signaling constraint of special relativity.
Here, we show that two parties with access to the primitive of non-local boxes
as described above are indeed able to perform
unconditionally secure bit commitment (BC) as well as one-out-of-two oblivious
transfer (1-2 OT).

A bit commitment protocol allows Alice and Bob to perform the
following task: Alice has chosen a bit $b$, and wants to convince
Bob that her choice is made without revealing the actual value of
$b$. Since Bob is inherently mistrustful, Alice sends him some
piece of evidence that she made up her mind. However, Bob still
has insufficient information to obtain $b$. Later on, Alice tells
Bob her choice $b'$ and Bob verifies that Alice is honest ($b'=b$)
using the piece of evidence from Alice. The problem of oblivious transfer was
introduced by~\cite{rabin:ot}. The variant of 1-2 OT
first appeared in a paper by Even, Goldreich and
Lempel~\citep{even:firstOT} and also, under a different name, in
the well-known paper by~\cite{wiesner:conjugate}. 1-2 OT
allows Alice and Bob to solve a seemingly uninteresting problem:
Alice has two bits $s_0$ and $s_1$. Bob wants to learn one of
them, but does not want to disclose to Alice which bit he is
interested in. However, Bob should also be restricted to learning
only one of Alice's inputs. It turns out that given 1-2 OT we can
perform any kind of two-party secure
computation~\citep{kilian:foundingOnOT}.

It has been understood for a long time that noisy channels
and preshared noisy correlations are sufficient to implement
secure two-party computations, via 1-2 OT.
~\cite{kilian:cryptogates} has
shown that noisy ``cryptogates'' (primitives with inputs and
outputs for each of the two players) can generically be used
to implement 1-2 OT. Based on the techniques of that paper
one would expect that non-local boxes would permit 1-2 OT,
since they provide some intrinsic noise.
This is indeed the case, but for more subtle reasons, as we shall discuss 
in the present paper.

We would also like to draw the reader's attention to the work of~\cite{wim:nonlocal,wim:thesis}, 
who shows that access
to perfect non-local boxes allows Alice and Bob to perform any
kind of distributed computation by transmitting only a single bit
of information. This is even true for slightly less perfect boxes
achieving weaker correlations~\citep{falk:nonlocal}.

\subsection{Related Work}\label{wolf}

Recently~\cite{wolf:otNL} suggested
           that 1-2 OT can be constructed using
           one non-local box alone. However, their version of 1-2 OT
           implicitly assumes that the non-local box acts as a kind of
           cryptogate: either the box has to wait until both players
           provide their input before it produces output, or its use is
           timed in the sense that the protocol will demand an input at
           a certain moment, and if a player does not supply one,
           uses a standard input instead (say, 0). Notice that the first
           possibility runs somewhat counter to the spirit 
           of non-local boxes, as it would allow signaling by
           delaying or not delaying an input. Non-local boxes, however,
           cannot be used to signal.
           That this assumption of synchronous input/usage of the box is
           vital to the result of Wolf and Wullschleger can easily be seen: 
           without this assumption, Bob can delay his choice of
           the selection bit indefinitely by simply deferring his use of
           the non-local box. This makes an important difference in
           reductions to 1-2 OT. Consider for example the standard
           reduction of OT to 1-2 OT (see Section $2(b)$ for
           definitions): The sender uses inputs $s_k = b$ and
           $s_{\bar{k}} = 0$ with $k \in_R \01$. The receiver uses input
           $c \in \01$. The players now perform 1-2 OT$(s_0,s_1)(c)$ after
           which the receiver holds $s_c$. Then the sender announces
           $k$. If $k = c$, the receiver succeeds in retrieving $b$ and
           otherwise he learns nothing. This happens with probability
           $p = 1/2$ and thus we have constructed OT from one instance
           of 1-2 OT. Clearly, this reduction fails if we use 1-2 OT
           based on the type of boxes suggested in~\citep{wolf:otNL}.
           The receiver simply waits for the announcement of $k$ to
           retrieve $b$ with probability $p = 1$. This was noticed independently
           by~\cite*{gisin:inprep}. However, the protocol of Wolf and Wullschleger
           forms a useful basis for our construction of 1-2 OT in Section~\ref{ot}.

\subsection{This Work}

Here, we demonstrate how to circumvent the problem of delay and construct 
a protocol for bit commitment and 1-2 OT based on non-local boxes.
This shows that superstrong non-local correlations in the form of non-local boxes
enable us to solve cryptographic problems otherwise known to
be impossible. Our work therefore creates a link between cryptographic problems 
and the nature of non-locality. In particular, our result implies that the no-signaling
principle and secure computation are compatible in principle.

\subsection{Outline}
Notation and definitions are introduced in Section~\ref{prelim}.
Section~\ref{bc} presents a protocol
for bit commitment based on non-local boxes. Finally, in Section~\ref{ot}, 
we show how to obtain 1-2 OT using
the same type of boxes.

\section{Preliminaries}\label{prelim}

\subsection{Notation}
Throughout this text, we say ``Alice \emph{picks $x$}'' if Alice chooses $x$ independently at random from the uniform distribution over all strings of a certain length.
We write $[n]$ for  $\{1,\ldots,n\}$, and $y \in_R S$ if $y$ is chosen uniformly at random from $S$.
In addition, we use $x \cdot y$ to denote the inner product $\sum_{i=1,\dots,n}x_i\cdot y_i \mod 2$ between strings $x=x_1 \dots x_n$ and $y=y_1\dots y_n$ from $\01^n$.
Furthermore, for strings $x \in \{00,01,10,11\}^\ast$ we
define $|\cdot|_{11}$ recursively: For the empty word $\epsilon$
define $|\epsilon|_{11}=0$. For $a,b \in \{0,1\}$ and strings
$x\in \{00,01,10,11\}^\ast$ define $|abx|_{11}=|x|_{11}+1$ if
$ab=11$ and $|abx|_{11}=|x|_{11}$ otherwise. Informally, for strings $x$ of even length,
$|x|_{11}$ is the number of substrings ``$11$'' in $x$ starting at
an odd position.

\subsection{Model and Definitions}\label{defs}

Throughout this text, we call the participant in a protocol \emph{honest} if he follows the
protocol. Since we are only interested in the case of unconditional security, a \emph{dishonest}
participant is not restricted in any way. In particular, he may lie about his own input, deviate from the protocol, or even abort the protocol completely.

\subsubsection{Non-local Boxes}

A non-local box (NL Box), sometimes also referred to as Popescu-Rohrlich box,
can be seen as a two-party primitive, generating correlations~\citep{popescu:nonlocal}.

\begin{definition}
A \emph{non-local box (NL Box)} is a two-party primitive between Alice and Bob,
in which Alice can input a bit $x \in \01$ and obtains an outcome $a \in \01$ and
Bob can input $y \in \01$ and obtains outcome $b \in \01$ such
that the following holds:
\begin{itemize}
\item Once Alice inputs $x \in \01$, she instantaneously receives outcome $a \in \01$,
\item Once Bob inputs $y \in \01$, he instantaneously receives outcome $b \in \01$,
\end{itemize}
such that $x \cdot y = a \oplus b$.
Further, we demand that for all $c_1,c_2,c_3 \in \01$ 
\[Pr[a=c_1|x=c_2,y=c_3] =Pr[b=c_1|x=c_2,y=c_3]=1/2.\]
\end{definition}
Observe that the last condition implies that these boxes 
cannot be used to signal,
because the outcome of $a$ is independent of $x$ and $y$ and also $b$ is independent of $x$ and $y$. 
It is worth mentioning that specifying the statistics of
           the primitive as we did, and disregarding the fact that
           outputs are obtained immediately after giving a local input,
           a non-local box is simply a special bidirectional channel, as
           proposed by~\cite{shannon:channels}. Of course, in general
           such channels cannot give an output immediately without
           having both inputs; non-local boxes \emph{can}, because they
           have no signaling capacity. Observe furthermore that the
           behaviour described in the definition parallels quantum
           mechanical experiments on entangled states: the outcomes are
           correlated in a way reflecting the measurement settings, but
           each experimenter obtains his outputs immediately.

Note that both Alice and Bob can wait indefinitely before providing their input to the NL Box.
Once they use the box, however, they will only obtain an outcome in accordance with the condition
given above. 
We say that Alice or Bob \emph{delay} the use of their box, if they wait longer than a given protocol
dictates before providing their input to the NL Box.

\subsubsection{Bit Commitment}

Bit commitment is a well known cryptographic primitive that plays an important role in many
other cryptographic protocols. It is defined as follows:

\begin{definition}
\label{DefComm}
\emph{Bit commitment} \emph{(BC)} is a two-party protocol between Alice (the committer) and Bob (the verifier), which
consists of two stages, the committing and the revealing stage, and a final declaration stage in which
Bob declares ``accept'' or ``reject''. The following requirements should hold:
\begin{itemize}
\item (Correctness) If both Alice and Bob are honest, then before the committing stage Alice decides on a bit $c$.
Alice's protocol depends on $c$ and any randomness used. At the revealing
stage, Alice reveals to Bob the committed bit $c$. Bob accepts.
\item (Binding) Assume (a possibly dishonest) Alice wants to reveal bit $c'$. Then always
\begin{eqnarray*}
&&\Pr[\mbox{Bob accepts }|\mbox{ Alice reveals }c'=0] + \\
&&\Pr[\mbox{Bob accepts } |\mbox{ Alice reveals }c'=1] \leq 1.
\end{eqnarray*}
\item (Concealing) If Alice is honest, Bob does not learn anything about $c$ before the revealing stage.
\end{itemize}
\end{definition}
We say that \emph{Alice cheats} if she chooses a bit $c'$ only after the committing stage and tries to get
Bob to accept $c'$ during the revealing stage. We also say that \emph{Alice cheats successfully}, if
Bob accepts the chosen $c'$. Furthermore, we say that \emph{Bob cheats} if he tries to obtain $c$ before
the revealing stage. \emph{Bob cheats successfully} if he obtains the correct $c$ before the revealing stage.
Note that our protocol for bit commitment is probabilistic and thus achieves statistical security for a
security parameter $n$. The sum of acceptance probabilities in the binding condition only needs to be smaller than $1 + \eps^n$ for some $0 \leq \eps < 1$. Likewise, the probability that Bob correctly
guesses bit $c$ before the revealing stage is $p \leq 1/2 + (\eps')^n$ for some $0 \leq \eps' < 1$.
By choosing $n$ large we can get arbitrarily close to the ideal scenario.

\subsubsection{Oblivious Transfer}

Different versions of oblivious transfer exist in the literature. Here, we will be concerned with one of the
most simple forms of oblivious transfer, namely 1-2 OT.

\begin{definition}
\emph{One-out-of-two oblivious transfer} \emph{(1-2 OT$(s_0,s_1)(c)$)} is a two-party protocol between Alice (the sender)
and Bob (the receiver), such that the following holds:
\begin{itemize}
\item (Correctness) If both Alice and Bob are honest, the protocol depends on Alice's two input bits $s_0,s_1 \in \01$
and Bob's input bit $c \in \01$. At the end of the protocol Bob knows $s_c$.
\item (Security against Alice) If Bob is honest, Alice does not learn $c$.
\item (Security against Bob) If Alice is honest, Bob does not learn anything about $s_{\bar{c}}$.
\end{itemize}
\end{definition}
Again, our protocol is probabilistic and achieves statistical security for a
security parameter $n$.
The probability that Bob learns $s_{\bar{c}}$ is $p \leq \eps^n$ for
some $\eps < 1$. Similarly, the probability that Alice correctly guesses $c$ is upper bounded by
$1/2 + (\eps')^n$ for some $0 \leq \eps' < 1$.

As we saw in Section $1(a)$, the fact that Alice and Bob can wait before using an NL Box can have an
effect on cryptographic reductions. In our example, we made use of the most simple form of oblivious transfer, 
i.e. an erasure channel.
\begin{definition}
\emph{Oblivious transfer (OT)} is a two-party protocol between Alice (the sender) and Bob (the receiver),
such that the following holds:
\begin{itemize}
\item (Correctness) If both Alice and Bob are honest, the protocol depends on Alice's input bit $b \in \01$.
At the end of the protocol, Bob obtains $b$ with probability $1/2$ and knows whether he obtained $b$ or not.
\item (Security against Alice) If Bob is honest, Alice does not learn whether Bob obtained $b$.
\item (Security against Bob) If Alice is honest, Bob's probability of learning bit $b$ does not exceed $1/2$.
\end{itemize}
\end{definition}

\section{BC from NL Boxes}\label{bc}
We now give a bit commitment
protocol based on NL Boxes.
Our protocol consists of $k$ blocks. In each block the parties use $2n+1$ shared non-local boxes. We later fix the security parameter $n$ such that 
we achieve sufficient security against Bob.

\begin{protocol}{1-NLBC(c) \quad  One Block}{}
\label{protOneRound}
\item[1-commit(c)]~
\begin{itemize}
\item Alice wants to commit to bit $c$. She encodes $c$ into a string $x$: She chooses $x \in \{0,1\}^\tno$ by randomly choosing the first $2n$ bits and then choosing $x_\tno \in \{0,1\}$ such that  $|x_1\dots x_{2n}|_{11}+x_{2n+1}+c$ is even.
\item Alice puts the bits $x_1,x_2,\dots,x_\tno$ into the boxes $1,2,\dots,\tno$. Let $a_1,a_2,\dots,a_\tno$ be Alice's output bits from the boxes. 
\item Alice computes the parity of all these output bits $A = \oplus_{i=0}^{\tno} a_i$ and sends $A$ to Bob.
\item Bob randomly chooses a string $y \in_R \{0,1\}^\tno$ and puts the bits $y_1,y_2,\dots,y_\tno$ into his boxes. We call the output bits from his boxes $b_1,b_2,\dots,b_\tno$.
\end{itemize}

\item[1-reveal(c)]~
\begin{itemize}
\item Alice sends $c$, her string $x$ and all her $\tno$ output bits to Bob. 
\item Bob checks if Alice's data is consistent: $\forall i \in \01^{\tno}, x_i \cdot y_i = a_i \oplus b_i$ and
$|x_1\dots x_{2n}|_{11}+x_{2n+1}+c$ is even. If not, he accuses her of cheating.
\end{itemize}
\end{protocol}

Define $C(x)$ to be the bit which is encoded by $x$. If Alice is honest, $C(x) = c$. It will be clear 
from our analysis in
Section $3(a)$, that if Alice cheats in one block of the
protocol Bob will notice this in the revealing stage with probability
$1/4$. To increase this probability, we can run many rounds of this protocol.

\begin{protocol}{NLBC(c) \quad The Full Protocol}{}
\item[commit(c)]~\\ Alice wants to commit to bit $c$. Then Alice and Bob run $k$ times 1-commit($c$) of 1-NLBC($c$).
\item[reveal(c)]~\\ Alice and Bob run $k$ times 1-reveal($c$) of 1-NLBC($c$).
\end{protocol}
\noindent

If Alice and Bob run the full protocol NLBC($c$) with $k$ rounds,
then the probability that Bob catches a cheating Alice becomes
larger. In fact, in a $k$ block protocol, the probability that Alice can cheat successfully is 
$\leq(3/4)^k$.
Even though Bob learns a little bit about the committed bit $c$ in
each block, we show below that the amount of information he learns
about $c$ can be made arbitrarily small.

\subsection{Security against Alice} \label{secSecurityAlice} 
Let us first analyze the security against a
cheating Alice for one block only. We show that no
matter which cheating strategy Alice uses, she is always detected
with probability at least $1/4$. There are two cases for Alice's
cheating strategy:
\begin{enumerate}
\item \emph{She has input something into all her boxes after the committing stage.} If she wants to reveal a bit different from $C(x)$ (for the
originally chosen $x$), she needs to change at least one of her
$x_i$. If she does not change the corresponding output bit $a_i$
and if Bob had input $y_i=1$ she will be caught. Similarly, if she changes $a_i$ but Bob had input $y_i=0$
she will be caught. Because $\Pr[y_i=1]=\Pr[y_i=0]=1/2$ she is
detected with probability at least $1/2$.

\item \emph{Alice delays her input to some boxes after the committing stage.} 
Without loss of generality
we can assume that all boxes have inputs before the revealing stage.
Otherwise, Alice's strategy
is equivalent to giving a random input and disregarding
both input and output. 
 
Suppose Alice sends bit $A'$ to Bob in the committing stage, pretending it was the parity of her $a_i$'s. She now wants to reveal.
Since the outputs of her delayed boxes are completely random to her, with
probability $1/2$, the parity of all $a_i$'s will be different from
$A'$. Thus, in this case she has to change at least one $a_i$. But
if $y_i=0$ ($y_i=1$) and Alice does (does not)  change $x_i$, she
is caught. Thus, Alice's cheating is detected with probability at least
$1/4$.
We now show that there is a cheating strategy for Alice,
such that she is only detected with probability $1/4$, if at least
$3$ boxes are used per block:
She first sends a random bit
$A$ to Bob in the committing stage and does not input anything into
her  boxes. In the revealing stage she chooses $x \in
\{00\}\{0,1\}^{2n-1}$ with $C(x)=c'$, where $c'$ is the bit she
wants to reveal. Then she puts the $x_i$'s into her boxes. With
probability $1/2$ the parity of the outputs $a_i$ from the boxes
is equal to $A$. Then she is lucky and proceeds with the protocol
as she was supposed to. If not she flips the bits $x_{1}$ and
$a_{1}$ and then goes on with the protocol as normal. Now, the
parity of the output bits is indeed equal to the $A$ she sent
before and the $x$-string still encodes $c'$. The changes are
detected by Bob iff $y_{1}=0$. If Bob is honest we have
$\Pr[y_{1}=0]=1/2$. Thus, a cheating Alice, using the above
strategy, is detected by an honest Bob with probability $1/4$.
\end{enumerate}
\noindent
Now, assume that Alice and Bob run a $k$-block protocol. 
Note that Alice may employ a different strategy (1 or 2, see above) for each
block. 
Assume that Alice employs strategy 2 in $k_*$ blocks and that she employs strategy 1 in $k-k_*$ blocks.
For strategy 1: She commits to $1$ in $k_1$ blocks and to $0$ in $k_0=k-k_*-k_1$ blocks, where $k_1 \leq k - k_*$.
Then security against Alice as in Definition \ref{DefComm} follows by proving that the following is close to $1$:
\begin{eqnarray*}
\Pr[\mbox{Bob accepts }|\mbox{ Alice reveals }c'=0] &+& \Pr[\mbox{Bob accepts } |\mbox{ Alice reveals }c'=1] \\
&\leq & (1/2)^{k_1}(3/4)^{k_*}+(1/2)^{k_0}(3/4)^{k_*}\\
\label{lastline} &=&(3/4)^{k_*}(1/2)^{k_0}\left(1+(1/2)^{k_1-k_0} \right).
\end{eqnarray*}
Without loss of generality
we can assume $k_1 \geq k_0$. Then $1+(1/2)^{k_1-k_0}\leq 2$. Thus, if $k_*> 2$ or $k_0 > 0$ the 
last expression is certainly less or equal to $1$. For $k_* \leq 2$ and $k_0 = 0$ the expression is 
upper bounded by $1+(1/2)^{k-2}$, which is in accordance with Definition \ref{DefComm}.

\subsection{Security against Bob}
\label{secSecurityBob}
Let us now analyze the security against a
cheating Bob. We first want to prove that Bob cannot
learn too much in one block. Bob can base his guess for $c$ on the
output of his boxes and the bit $A$ he receives from Alice. Note
that after Bob has received the bit $A$, he learns the inner
product of $x$ and $y$, because: $x \cdot y =\bigoplus_{i=1}^\tno
x_i \cdot y_i =\bigoplus_{i=1}^\tno a_i\oplus b_i=A \oplus
\bigoplus_{i=1}^\tno b_i$.

We want to argue that this is all Bob learns about $x$ (and
therefore $c$). In the trivial case $y=0^\tno$ it is easy to see
that Bob learns nothing, because his output bits $b_i$ are
uniformly random and the bit $A$ he receives does not contain any
information since $A = \bigoplus_i^\tno b_i$. For that reason we
will not consider the case $y =0^\tno$ in our further analysis.

Assume now Bob chooses $y \in \{0,1\}^\tno \backslash \{0\}^\tno$.
Furthermore, assume that Alice and Bob follow the above protocol,
but this time Alice does not commit to a bit but rather chooses a
uniformly random string $x \in \{0,1\}^\tno$ . First note that as above Bob still
learns $x\cdot y$. Since $|\{x: x \cdot y =1\}|=|\{x: x \cdot y =0\}|$, $x \cdot y $
contains exactly one bit of information about $x$. But also,
since the boxes are non-signaling and Alice only sends one bit,
Bob can learn at most one bit of information about $x$. Therefore,
the only thing Bob learns about $x$ is $x \cdot y $. Since in this
changed protocol Bob learns precisely $x \cdot y $, also in the original
protocol Bob learns precisely $x \cdot y $ and nothing else.

The following lemma can be used to upper bound Bob's information
gain in one block, by proving that $x \cdot y$ (Bob's only information about Alice's commitment) is always almost uniformly  distributed.
\begin{lemma}
\label{lemmaBoboneblock} Assume Alice and Bob execute one block of the
protocol with $\tno$ NL Boxes, where Bob chooses some
$y\in\{0,1\}^\tno\backslash\{0\}^\tno$ and Alice commits to some $c\in
\{0,1\}$. Then the probability for $x \cdot y =c$, averaged over all $x
\in C^{-1}(c)$, obeys
$$\left|\Pr_{x,C(x)=c}[x \cdot y =c]- 1/2\right|\leq 1/2^{n+1}.$$
\end{lemma}
\begin{proof}
We write $p^c_y$ as a shorthand for $\Pr_{x,C(x)=c}[x \cdot y =c]$.
The proof is by induction on $n$. For $n=0$ the statement is
easily seen to be true. Assume now $n>0$. Let $y_1,y_2$ be the
first two bits of $y$ and $y'$ the rest, i.e. $y=y_1y_2y'$. To
explain the argument, let us for instance look at the case
$y_1y_2=01$. For any $x' \in \{0,1\}^{2n-1}$ we have $C(x')\oplus
(x_1\cdot y_1) \oplus (x_2 \cdot y_2)=C(x_1x_2x')$ if $x_1x_2\in \{00,10,11\}$ and we have
$C(x')\oplus (x_1\cdot y_1) \oplus (x_2 \cdot y_2) =\overline{ C}(x_1x_2x')$ if
$x_1x_2=01$. This observation yields
$$
p^c_{01y'}=\Pr[x_1x_2\in \{00,10,11\}]p^c_{y'}+\Pr[x_1x_2=01]p^{\bar c}_{y'}
     =1/2+1/4(p^c_{y'}-p^{\bar c}_{y'}),
$$
where we used in the second equality $\Pr[x_1x_2=x_1'x_2']=1/4$ for
any $x_1'x_2'$ and $p^c_{y'}+p^{\bar c}_{y'}=1$ for $y'\neq
0^{2n-1}$. By the inductive assumption $\left|p^c_{y'}-p^{\bar
c}_{y'}  \right|\leq 2^{-n+1}$ and thus $|p^c_{01y'}-1/2|\leq
2^{-(n+1)}$.
In the other cases for $y_1y_2$ we get
\begin{eqnarray*}
p^c_{00y'}&=&\Pr[x_1x_2\in \{00,10,01\}]p^c_{y'}+\Pr[x_1x_2=11]p^{\bar c}_{y'}\\
p^c_{10y'}&=&\Pr[x_1x_2\in \{00,01,11\}]p^c_{y'}+\Pr[x_1x_2=10]p^{\bar c}_{y'}\\
p^c_{11y'}&=&\Pr[x_1x_2=00]p^c_{y'}+\Pr[x_1x_2\in\{01,10,11\}]p^{\bar c}_{y'},
\end{eqnarray*}
from which the bound follows by the same argument as above.
\end{proof}
\noindent
We now analyze a $k$-block protocol, where for simplicity $k$
is even. We only consider the case where Alice commits to $c=0$
and $c=1$ each with probability $1/2$.
\begin{lemma}
\label{lemmaBob} Assume Alice and Bob run a $k$-block protocol in
which in each block $\tno$ boxes are used. Then the probability
that Bob can guess the committed bit correctly is upper bounded by
$1/2 + k/2^{n+1}$.
\end{lemma}
\begin{proof}
Let $r_i$ be Bob's best guess for $c$ using only $x \cdot y $ from the
$i$-th block. Set $\epsilon_i$ such that $1/2 +\epsilon_i=
\Pr[c=r_i]$. By Lemma~\ref{lemmaBoboneblock}, $0 \leq \epsilon \leq
1/2^{n+1}$. Note that Bob's only information about $c$ is
$r_1,\dots ,r_k$.

Let us think of the process of how $r_i$ is obtained in a way
which is easier to analyze but equivalent to the original: With
probability $1-2\epsilon_i$ (a) the bit $r_i$ is chosen randomly
from $\{0,1\}$ and with probability $2\epsilon_i$ (b) $r_i$ is set
to $c$.

By the union bound the probability that at least once during the
$k$ blocks case (b) occurs is at most $\sum_{i=1}^k
2\epsilon_i\leq k/ 2^{n}$. Thus, with probability at least
$1-k/2^{n}$ the bits $r_1,\dots,r_k$ are completely random.
Bob's probability of guessing correctly is upper bounded by
$1/2(1-k/2^{n})+k/2^{n}=1/2 + k/2^{n+1}$.
\end{proof}
\noindent
Note that the analysis in Lemma \ref{lemmaBob} is not tight,
but sufficient for our purposes.

\section{1-2 OT from NL Boxes}\label{ot}

We now show how to construct 1-2 OT from NL Boxes. We thereby
assume that Alice and Bob have access to a secure bit commitment
scheme BC as given in Section~\ref{bc} for sufficiently large $k$.
Our protocol extends the protocol suggested by~\cite{wolf:otNL} and uses an idea presented in the
context of quantum oblivious transfer by~\cite{crepeau:reductions, crepeau:qot}
and~\cite{crepeau:weakened}.

\subsection{Protocol}

Before presenting the actual protocol, we briefly discuss the
intuition behind it. The rough idea is that using NL Boxes, we can
approximate an erasure channel from Alice and Bob: Suppose Alice
has input $v \in \01$. She picks $y \in_R \01$, sets $r_{y} = v$
and picks $r_{\bar{y}} \in_R \01$. If Alice inputs $x = r_0 \oplus
r_1$ and Bob inputs $y' \in_R \01$ to an NL Box they will obtain
outputs $a$ and $b$ with $a \oplus b = x\cdot y'$. If Alice now
sends $m = r_0 \oplus a$ to Bob, Bob will obtain $r_{y'}$ by
computing $m \oplus b =  r_0 \oplus a \oplus b = r_0 \oplus (r_0
\oplus r_1) y' = r_{y'}$. He cannot obtain more than one bit of
information, as he receives only one bit of communication from
Alice. Now Alice announces $y$ to Bob. If $y = y'$, Bob received
Alice's input bit $r_{y'} = v$. This happens with probability $1/2$.
The only trick we need, is to make sure Bob actually did use the
NL Box and made his choice of $y'$ before Alice's  announcement. To
achieve this, bit commitment is used in step 2.

\begin{protocol}{1-2 NLOT($s_0,s_1$)($c$)}{}
\item For $1 \leq i \leq 2n$:
\begin{itemize}
\item Alice picks $r_{0,i},r_{1,i} \in_R \01$.
\item Bob picks $y_i' \in_R \01$.
\item Alice and Bob use one NL Box with inputs $x_i = r_{0,i} \oplus r_{1,i}$ and
$y_i'$ respectively. Alice gets $a_i$, Bob $b_i$.
\end{itemize}
\item  For $1 \leq i \leq n$:
\begin{itemize}
\item Alice and Bob run commit($y_i'$),commit($b_i$),commit($y_{i+n}'$),commit($b_{i+n}$), where Bob is the sender.
\item Alice picks $k_i \in_R \01$, and announces it to Bob.
\item Alice and Bob run reveal($y_{i+k_i n}'$) and reveal($b_{i+k_i n}'$), where Bob is the sender.
\item Alice checks that $x_{i+k_i n} \cdot y_{i+k_i n}' = a_{i+k_i n} \oplus b_{i+k_i n}$ and otherwise
aborts the protocol.
\item Alice sets $r_{0,i} \leftarrow r_{0,i+\bar{k}_i n}$, $r_{1,i} \leftarrow r_{1,i+\bar{k}_i n}$ and $a_{i} \leftarrow a_{i+\bar{k}_i n}$.
Bob sets $b_{i} \leftarrow b_{i+\bar{k}_i n}$ and $y_{i}' \leftarrow y_{i+\bar{k}_i n}'$.
\end{itemize}
\item For $1 \leq i \leq n$:
\begin{itemize}
\item Alice sends $m_i = r_{0,i} \oplus a_i$ to Bob.
\item Bob computes $v'_i = m_i \oplus b_i = r_{y_i',i}$.
\item Alice picks $y_i \in_R \01$, sets $v_i = r_{y_i,i}$ and announces $y_i$ to Bob.
\end{itemize}
\item Bob picks $J_0, J_1 \subset [n]$, subject to $|J_0| = |J_1| = n/3$, $J_0 \cap J_1 = \emptyset$
and $\forall i \in J_c$, $y_i = y'_i$. He announces $J_0$, $J_1$ to Alice.
\item Alice receives $J_0$, $J_1$, checks that $J_0 \cap J_1 = \emptyset$ and otherwise
aborts the protocol. She computes $\hat{s}_0 = s_0 \oplus \bigoplus_{j \in J_0} v_j$
and $\hat{s}_1 = s_1 \oplus \bigoplus_{j \in J_1} v_j$. She announces
$\hat{s}_0$,$\hat{s}_1$ to Bob.
\item Bob now computes $s_c = \hat{s}_c \oplus \bigoplus_{i \in J_c} v'_i$.
\end{protocol}

\subsection{Correctness}

We first need to show that if both parties are honest, Bob succeeds in retrieving $s_c$ with
high probability. Note that Bob can retrieve $s_c$, if he can construct a set $J_c \subset [n]$ with
$|J_c| =n/3$ where $\forall i \in J_c, y_i = y'_i$, since only then $\forall i \in J_c, v_i = v'_i$ and
he can compute
$$
\hat{s}_c \oplus \bigoplus_{i \in J_c} v'_i = s_0 \oplus \bigoplus_{j \in J_c}
v_j \oplus \bigoplus_{i \in J_c} v'_j = s_c.
$$
We are thus interested in the probability of Bob constructing such a set successfully.
Let $X_i$ be the random variable such that $X_i = y_i \oplus y'_i$. Note that since Alice and Bob
choose $y_i$ and $y'_i$ independently uniformly at random, the random variable
$S_n = \sum_{i = 1}^n X_i$ is binomially distributed. From Hoeffding's inequality~\citep{hoeffding:bound} we obtain
\begin{equation}\label{hoeffding}
\Pr\left(S_n - \frac{n}{2} \geq \eps\right) \leq e^{-\frac{2\eps^2}{n}}.
\end{equation}
Then,
\begin{eqnarray*}
\Pr(\mbox{Bob gets }s_c) &=& \Pr\left(\#\{i|y_i = y'_i\} \geq \frac{n}{3}\right) \\
&=&1 - \Pr\left(\#\{i|y_i = y'_i\} < \frac{n}{3}\right)\\
&=&1 - \Pr\left(S_n > \frac{2n}{3}\right)\\
&\geq&1 - \Pr\left(S_n - \frac{n}{2} \geq \frac{n}{6}\right)\\
&\geq&1 - e^{-\frac{n}{18}},
\end{eqnarray*}
where the last inequality comes from equation~(\ref{hoeffding}).
Thus the probability of Bob failing is exponentially small in $n$.

\subsection{Security against Alice}
Suppose that Bob is honest, but Alice tries to learn $c$.
As outlined in Section~\ref{prelim}, NL Boxes do not allow signaling and therefore Alice learns
nothing during step 1 of the protocol. Due to the concealing properties of the bit commitment
scheme, Alice's information gain in step 2 is negligible for a sufficiently large security parameter $k$.
Thus the only time she receives information from Bob
is during step 4. Note that Bob picks $y_i'$ independently of $y_i$. Alice has
no information on $y'_i$. This means
that the elements of the sets $J_0$ and $J_1$ are independent of $c$ from
Alice's point of view: their composition depends only on whether $y'_i = y_i$ for a given $i$.
Alice thus learns nothing from observing the sets $J_0$ and $J_1$.

Note that Alice gains nothing from trying to delay her own boxes:
By delaying boxes in the commitment protocol employed in step 2, she
will only remain more ignorant about Bob's commitment. Furthermore, 
each round $i$ 
in step 1 corresponds to Alice using an erasure channel with input
$v_i = r_{y_i,i}$, because the following two conditions are satisfied: 
step 2 ensures us that Bob uses this channel, and, since Alice 
sends $y_i$ to Bob during step 3, Bob knows whether he obtained $v_i$.
The situation where Alice delays using the boxes is equivalent to
using the channel with a randomly chosen input and gives her no additional
advantage.
Since we can construct an erasure channel, we thus obtain an 1-2 OT via
the above construction~\citep{crepeau:reductions}.

\subsection{Security against Bob}
Now suppose that Alice is honest, but Bob tries to learn more than $s_c$.
We now show that Bob can retrieve exactly one of the bits $s_0,s_1$. In particular,
we show that he cannot compute any function $f$ of $s_0$ and $s_1$ 
which depends on both input bits.\footnote{A function $f$ depends on the $j$-th input argument if there is an input to $f$ such that changing the $j$-th argument changes the value of $f$.}

Because Alice is honest, all $v_i$ are
independent. Furthermore, since the sets $J_0$ and $J_1$ are disjoint,
it follows that $r = \oplus_{j \in J_0} v_j$ and $r' = \oplus_{j \in J_1} v_j$
are independent. All Bob receives from Alice is $\hat{s}_0 = s_0 + r$ and
$\hat{s}_1 = s_1 + r'$. Thus, in order to compute any function $f$ of $s_0,s_1$ which depends on 
both input bits, Bob needs to learn both $r$ and $r'$. Bob will only obtain $r$ and $r'$
and then also learn more than one of the bits $s_0,s_1$, if he succeeds in creating two sets $J_0,J_1 \subset [n]$
with $J_0 \cap J_1 = \emptyset$ and $|J_0| = |J_1| = n/3$ such that $\forall i \in J_0 \cup J_1, y_i = y'_i$.
We are therefore interested in the probability that Bob can successfully construct two such sets.

In order to construct such sets, Bob may try to delay using some of the NL Boxes during step 1. This
will enable him
to wait for the announcement in step 3, to force $y_i = y_i'$ and obtain $v_{y_i}$ with certainty.
By assumption, the bit commitment scheme is binding for sufficiently large $k$ and thus Bob cannot try to fool Alice by
breaking the commitment itself. However, he can try to commit to random values and escape detection during
step 2. In particular, he can choose to be honest in step 1 for exactly one NL Box in runs $i$ and $n+i$.
Without loss of generality, suppose he was honest in run $n+i$ and delayed use of the box in run $i$.
He then commits once to the outcome of the honest box, and once to $y_i' = 1$ and a random $b_i \in_R \01$.
The probability that Alice challenges him on the box he has been honest with in step 1 is $1/2$. Then
he has succeeded to cheat on one of the bits, $y_i'$, and will obtain $v_{y_i}$ with certainty.
However, with probability $1/2$ Alice will challenge him on the other NL Box.
In this case he can escape detection with probability $1/2$: He announces $y_i'$ and $b_i$ and hopes
that this matches
the input of Alice's box. He will have committed to the correct $b_i$ with probability 1/2 and then he escapes
detection.
Thus the total probability of cheating successfully on one of the bits is given by
$1/2 + (1/2)(1/2) = 3/4$. Let $C \subseteq [n]$ with $k = |C|$, $0 \leq k \leq n$ denote the
set of indices on which Bob tries to deceive Alice. He will remain undetected with probability
$$
\Pr(\mbox{Bob successfully cheats on }k\mbox{ bits}) = \left(\frac{3}{4}\right)^k.
$$
Suppose now, that Bob successfully cheated on $k$ bits. We are then interested in bounding
the probability of constructing two valid sets if Bob already has $k$ valid entries.
Note that we now only consider the probability of achieving $y_i = y'_i$ for indices $i \notin C$
and then $\#\{i|y_i = y'_i\} = (n-k) - S_{n-k}$. For $k < n$,
\begin{eqnarray*}
\Pr(\mbox{Bob gets }s_0\mbox{ and }s_1) &=& \left(\frac{3}{4}\right)^k \Pr\left(\#\{i|y_i = y'_i\} \geq \frac{2n}{3} - k\right)\\
&\leq&\left(\frac{3}{4}\right)^k \Pr\left(S_{n-k} \leq \frac{n}{3}\right)\\
&=& \left(\frac{3}{4}\right)^k \Pr\left(\frac{n-k}{2} - S_{n-k} \geq \frac{n-3k}{6}\right)\\
&\leq& \left(\frac{3}{4}\right)^k e^{-2\left(\frac{(n-3k)^2}{18(n-k)}\right)}
\end{eqnarray*}
If $k = n$, Bob will be caught with probability $(3/4)^n$.
Thus the probability of Bob deceiving Alice can be made arbitrarily small by choosing $n$ large.

\section{Conclusion}
We have shown how to obtain protocols for bit commitment and one-out-of-two oblivious transfer
given access to non-local
boxes. This creates a link between cryptographic problems,
which may appear very artificial, and non-local
correlations: If such NL Boxes
were available in nature, we could implement these cryptographic
protocols securely which is known to be impossible to achieve
using quantum mechanics alone.

Interestingly, the quantum mechanical impossibility proofs
for bit commitment and coin tossing~\citep{lo&chau:bitcom,lo&chau:bitcom2,mayers:trouble,mayers:bitcom,lo:insecurity} via the so-called
EPR-attack are the quantum version of delaying the input.
One may want to go back to explore why we could circumvent
this attack here, and the reason seems to be that the
NL Box is more like a quantum mechanical entangled
state \emph{together with an encasing experimental setup},
which enforces that the particles can only be measured
separately. In contrast, for the EPR-attack to work, Alice has
to be able to perform arbitrary collective operations on her
qubits. 

\section{Acknowledgments}
We thank the Newton Institute Cambridge for hosting the QIS
workshop where a part of this paper originated. This project was
supported by the EU under project RESQ (IST-2001-37559). MC
and AW acknowledge furthermore support by the
U.K.~Engineering and Physical Sciences Research Council.
MC acknowledges the support of a DAAD Doktorandenstipendium;
HB, FU and SW receive support from the NWO vici project 2004-2009.
\par\noindent
We would also like to thank Stefan Wolf and J\"urg
Wullschleger for discussions on their work~\citep{wolf:otNL}.
Furthermore we would like to thank Serge Fehr and Robbert de Haan
for useful discussions about 1-2 OT.

\label{lastpage}
\end{document}